\newcommand\astroph[2]{{\it `#1'} {\rm (see astro-ph/#2)}}
\newcommand\vpy[3]{{\bf #1}, #2 (#3).}
\newcommand\mnras[3]{{\rm Mon.~Not.~R.~Astron.~Soc.~\vpy{#1}{#2}{#3}}}

\documentstyle[prl,aps,epsf,floats,twocolumn]{revtex}

\begin{document}
\twocolumn[\hsize\textwidth\columnwidth\hsize\csname @twocolumnfalse\endcsname
\title{New method of extracting non-Gaussian signals}

\author{
  Jiun-Huei Proty Wu$^{1,2}$\footnote{jhpw@astro.berkeley.edu}
  }
\address{
$^1$Astronomy Department,
 University of California, Berkeley, CA 94720-3411, USA\\
$^2$Department of Applied Mathematics and Theoretical Physics,
University of Cambridge, CB3 9EW, UK}

\maketitle

\begin{abstract}

  We propose a new general method of extracting non-Gaussian features
  in a given field.
  It sets the mean of the field to zero and
  renormalizes its Fourier (or multipole) power to white noise,
  while keeping the phase information unchanged.
  Using simulated cosmic microwave background (CMB)
  as an example, 
  we demonstrate the power of this method
  under many challenging circumstances.
  In particular,
  we show its capability of detecting cosmic strings in the CMB.
\end{abstract}

\pacs{95.75.-z; 98.80.-k; 98.70.Vc; 11.27.+d}

]

\noindent
{\bf A.\ Introduction}:
Searching for and characterizing the non-Gaussianity (NG) of a given field
has been a vital task in many fields of science,
because we expect 
the consequences of different physical processes to carry
different statistical properties.
To illustrate this point,
we shall employ the cosmic microwave background (CMB) \cite{HuSugSil}
as an example, which features several prominent aspects of our interest.

With the cosmological principle as the basic premise,
there are currently two main competing theories 
for the origin of structure in the universe---inflation \cite{Guth}
and topological defects \cite{VilShe,Wu1}.
Although 
the beauty and simplicity of the former appears to have
enticed more adherents,
the latter can still coexist with the former.
In particular,
the observational verification of defects will
have certain impact to the grand unified theory,
because they are an inevitable consequence of 
the spontaneous symmetry-breaking phase transition in the early universe.
In addition to the conventional study of the power spectra
of cosmological perturbations,
another way to distinguish these models is via
the search for intrinsic NG in the perturbations---while
the standard inflation predicts Gaussianity,
theories like isocurvature inflation \cite{Pee}
and topological defects provide NG.

The cosmological test for NG can be implemented in both 
large-scale structure
(e.g.,~\cite{LSS-NG}) and CMB (e.g.,~\cite{Turok}).
The former is less favored because
its intrinsic statistical properties
can be easily distorted by the late-time non-linear gravitational interaction.
Even in the use of CMB,
there are still several observational and theoretical challenges.
For example,
the intrinsic statistical features of the CMB anisotropies
may be obscured by
the foreground contamination 
(from the radio emission of our own Galaxy, distant galaxies, etc.),
instrumental noise,
and sample variance;
the central limit theorem (CLT) will force
the large-scale perturbations to be Gaussian
even if 
very strong NG mechanisms operate on smaller scales.
Adding the fact that
high-precision and high-resolution observation is yet available,
various Gaussianity tests of the CMB data in the literature
have so far drawn no robust conclusion 
\cite{Bromley}.
Nevertheless,
a flood of high-accuracy data will be available in the near future
\cite{cmbexp,MA1,B98},
and we shall need a comprehensive formalism to deal with them.

In the literature,
the most commonly used statistics for the NG test are
moments, cumulants,
$n$-point correlation functions,
bispectra,
Minkowski functionals,
peak statistics, etc (for a review see, e.g.,~\cite{Bromley,WinitzkiWu}).
These statistics can be defined in different statistical spaces such as
the real, Fourier, wavelet, and eigen-spaces.
Here we shall combine both the real and Fourier spaces
to propose a new method
that helps extract out the NG from a given field.
With the extracted NG,
it then becomes easier for the conventional statistics to characterize
and then identify the NG originated from different physical processes.


\noindent
{\bf B.\ A new method}:
The defining property of a Gaussian field
is that the higher-order (greater than two) reduced moments vanish.
This means that
the only information that any Gaussian field carries is
its mean and
the two-point correlation function, or equivalently,
the power spectrum in the Fourier space.
Therefore we can refer to the mean and the power spectrum
as the `Gaussian components' of a field,
and any extra information will indicate NG.
Thus, the new method aims to nothing but
`removing' these Gaussian components.
Here we shall use an $n$-dimensional field $\Delta({\bf x})$
to demonstrate this formalism.

First
  we Fourier transform $\Delta({\bf x})$ to obtain $\widetilde{\Delta}({\bf k})
  =\int dx^n \Delta({\bf x}) e^{-i{\bf k}\cdot{\bf x}}$.
Second,
we estimate the power spectrum as
  $C_k=\langle|\widetilde{\Delta}({\bf k})|^2\rangle_k/V^n$,
  where $k\equiv |{\bf k}|$ and
 $V^n$ is the $n$-dimensional volume of the field.
Third,
  we set $\widetilde{\Delta}({\bf 0})=0$
  and define ($\forall {\bf k} \textrm{ with } C_k \neq 0$)
  \begin{equation}
  \label{Delta_p}
    \widetilde{\Delta}_P({\bf k})
    = \widetilde{\Delta}({\bf k})
     {{C_k^{-1/2}}}{P_k^{1/2}},
  \end{equation}
  where
  $P_k$ is a given function of $k$.
Finally,
  we inverse Fourier transform $\widetilde{\Delta}_P({\bf k})$
  back to the real space $  \Delta_P({\bf x})$.
Now the field $\Delta_P({\bf x})$ has a mean zero and a power spectrum
renormalized to $P_k$.
Therefore the original field can be written as
$
  \Delta({\bf x}) = \Delta_P({\bf x}) \otimes D({\bf x})\otimes Q({\bf x})
     +\overline{\Delta}
$,
where an overbar denotes the mean of a field,
the symbol $\otimes$ denotes a convolution,
and
$D({\bf x})$ and $Q({\bf x})$ are
the inverse Fourier transforms of the symmetric fields 
$\widetilde{D}({\bf k})\equiv {C_k^{1/2}}$
and 
$\widetilde{Q}({\bf k})\equiv {P_k^{-1/2}}$ respectively.
We shall investigate the simplest case where $P_k=1$,
and leave the discussion for other forms of $P_k$ later.
In this case,
the field $\Delta({\bf x})$ will be `whitened' in Fourier power
through the above procedure,
and we shall use a superscript `W' to denote such whitened fields.
In the real space, this means
\begin{equation}
  \label{Delta-decompose-real}
  \Delta({\bf x}) = \Delta^{\rm W}({\bf x}) \otimes D({\bf x})
     +\overline{\Delta}.
\end{equation}
Now
the Gaussian components, $\overline{\Delta}$ and $D$,
are separated from the rest.
Therefore if $\Delta$ is Gaussian,
then all samples in $\Delta^{\rm W}$ should appear 
as uncorrelated white noise.
Otherwise $\Delta^{\rm W}$ would contain all the NG features.

We note that 
the above new method (eq.~[\ref{Delta_p}]) is equivalent to the matrix manipulation
${\bf d}_P={\bf P}^{1/2} {\bf C}^{-1/2}{\bf d}$,
where ${\bf d}\equiv \Delta$, 
and ${\bf P}$ and ${\bf C}$ are the two-point correlation matrices
specified by $P_k$ and $C_k$ respectively.
In signal processing,
it is common to model ${\bf C}$ as a linear sum of the signal and noise,
i.e.\ ${\bf C}={\bf S}+{\bf N}$. Combined with the Wiener filtering
${\bf d}^{\rm WF}={\bf S} {\bf C}^{-1}{\bf d}$,
the whitening procedure is then 
${\bf d}^{\rm W}={\bf S}^{1/2} {\bf C}^{-1}{\bf d}$.
It 
not only removes the power on scales where the noise dominates,
but also equalizes the power on scales where the signal dominates.

To see how this formalism can work in practice,
we consider a linear sum of a Gaussian and a NG fields:
$\Delta = \Delta_{\rm (G)}+\Delta_{\rm (NG)}$.
Assuming no correlation exists between $\Delta_{\rm (G)}$ and $\Delta_{\rm (NG)}$,
as in normal situations,
we have $C_k=C_{k{\rm (G)}} + C_{k{\rm (NG)}}$, leading to
\begin{equation}
  \label{linear-sum-fourier}
  \widetilde{\Delta}^{\rm W} 
  = 
  \frac{
  {C_{k{\rm (G)}}^{1/2}}  \, \widetilde{\Delta}_{\rm (G)}^{\rm W} +
  {C_{k{\rm (NG)}}^{1/2}} \, \widetilde{\Delta}_{\rm (NG)}^{\rm W}
  }
  {\left[{C_{k{\rm (G)}} + C_{k{\rm (NG)}}}\right]^{1/2}}.
\end{equation}
Thus we see that if $C_{k{\rm (G)}} < C_{k{\rm (NG)}}$ at a certain $k$,
the NG part $\widetilde{\Delta}_{\rm (NG)}^{\rm W}$ will dominate.
As a consequence,
we expect the NG feature to show up in ${\Delta}^{\rm W}$
as long as the NG signal dominates within a certain range of $k$.
As we shall see later,
in most cases the NG features in $\Delta^{\rm W}$
are already largely visually recognizable.
Therefore, 
instead of trying further statistics to characterize our $\Delta^{\rm W}$,
we shall concentrate more on the use of this formalism,
in the face of various challenging situations.

\begin{figure}[t]
  \centering
  \leavevmode\epsfxsize=8cm \epsfbox{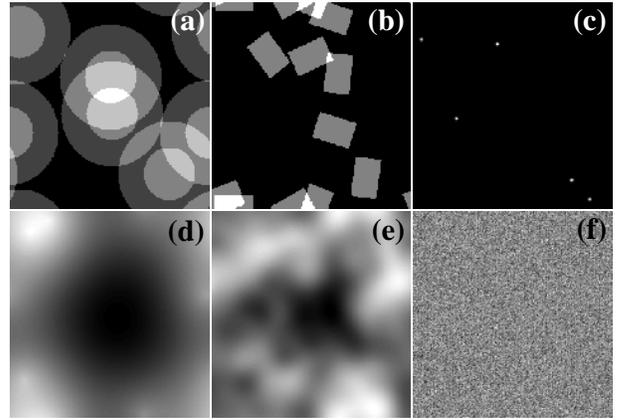}\\
  \caption
  {Six components in a simulated CMB map
    of $1^\circ \times 1^\circ$.
  }
  \label{fig-var-components}
\end{figure}

\begin{figure}[t]
  \centering
  \leavevmode\epsfxsize=8cm \epsfbox{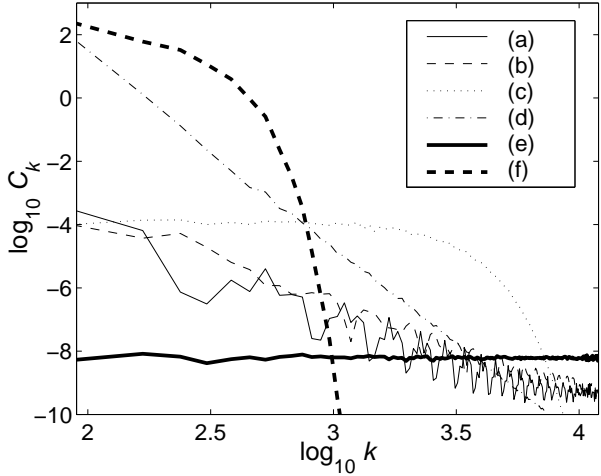}\\
  \caption
  {Power spectra of the six components in Fig.~\ref{fig-var-components}.
     }
  \label{fig-Pk_test2}
\end{figure}

\begin{figure}[t]
  \centering
  \leavevmode\epsfxsize=8cm \epsfbox{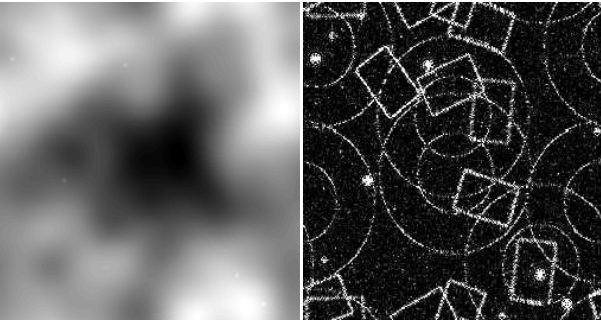}\\
  \caption
  {{\em Left}: a simulated CMB map $\Delta$, with the six components
    presented in Fig.~\ref{fig-var-components} linearly added up.
    {\em Right}: the square of the extracted NG signal $(\Delta^{\rm W})^2$.
    The color scheme in the left plot is between the minimum and the maximum
    while that in the right plot is from zero to one sigma square.
    }
  \label{fig-test2}
\end{figure}

\noindent
{\bf C.\ Against various NG components}:
We now apply this new formalism to the CMB anisotropies
$\Delta({\bf x}) \equiv \left[ T({\bf x})-\overline{T}\right]/\overline{T}$,
where $T({\bf x})$ is the CMB temperature at the direction ${\bf x}$.
In the `small-field limit' where
the curvature of the sky can be ignored,
the conventional multipole decomposition 
$\Delta({\bf x})=\sum_{\ell,m} a_{\ell m}Y_{\ell m}({\bf x})$
simplifies to a two-dimensional Fourier transform
such that
the multipole number is simply
$\ell\equiv |{\bf k}|=k$ and
the angular power spectrum becomes
$C_\ell=\langle |a_{\ell m}|^2\rangle\equiv C_k$.
We shall work in this small-field limit throughout the paper.
First we simulate a CMB field of
size $1^\circ\times 1^\circ$ and resolution $256^2$,
with six components mimicking results of different physical processes
(see Fig.~\ref{fig-var-components}):
(a) 10 filled circles of diameters 64 and 128 grids.
(b) 10 rectangles of size $32\times 48$ grid$^2$.
The sharp edges of components (a) and (b) are
similar to the integrated Sache-Wolf (ISW) effect
produced by cosmic strings \cite{KS}.
(c) 5 sharp circular points (like distant galaxies),
each with a power spectrum $C_{k{\rm (c)}}=\exp\left[-(rRk)^2\right]$,
where $R$ is the angular size of the field ($1^\circ$ here)
and $r=0.005$.
(d) 5 diffuse circular points (like cosmic defects such as textures \cite{TurSpe}),
each with a power spectrum $C_{k{\rm (d)}}=[1+(sRk)^t]^{-1}$
where $s=0.5$ and $t=6$.
(e) a Gaussian background (like the inflationary perturbations)
of power spectrum
$
  C_k \propto
  k^{-2} \exp{\left[-\left(\frac{k}{1200}\right)^2\right]}
$,
which fits well on small scales ($k \gg 1$) with
that of a standard CDM model
computed from CMBFAST \cite{cmbfast}.
Even in the cosmic-defect-dominated models,
this Gaussian background is still expected as the contribution 
from before the last scattering epoch $t_{\rm ls}$.
This is because, in any cases,
the sub-degree perturbations are smeared away
due to the photon diffusion damping by $t_{\rm ls}$ \cite{Silk}
(thus creating the exponentially decaying power as seen above),
while
the super-degree perturbations
must be nearly Gaussian due to causality and the CLT
(the angular size of the horizon at $t_{\rm ls}$
 is about one degree).
(f) Gaussian noise (like instrumental effects) of white power spectrum.
Thus the components (a)--(d) are non-Gaussian,
while (e) and (f) are Gaussian.
Fig.~\ref{fig-Pk_test2} shows the power spectra of each components.
These six components are then linearly added up
$\Delta=\sum_i \Delta_{i}$,
with RMS ratios (a--f)
$1:1:10:500:1000:0.2$ (see Fig.~\ref{fig-test2} left).
We then apply our new method to obtain the whitened field $\Delta^{\rm W}$
(see Fig.~\ref{fig-test2} right).

The above results indicate three things.
First,
although the NG components ($\Delta_{\rm a}$--$\Delta_{\rm d}$)
dominate the Gaussian ones ($\Delta_{\rm e}$ and $\Delta_{\rm f}$)
in power only within a certain range of $k$,
their NG features are almost fully recovered in $\Delta^{\rm W}$.
This confirms the expectation 
associated with equation (\ref{linear-sum-fourier}) 
and its context.
Second,
the centers of the diffuse points in $\Delta_{\rm d}$ 
are now fully recovered in $\Delta^{\rm W}$,
although hardly recognizable in $\Delta_{\rm d}$
(see Fig.~\ref{fig-var-components} (d)).
Finally,
although $\Delta_{\rm c}$ and $\Delta_{\rm d}$ are much stronger than
$\Delta_{\rm a}$ and $\Delta_{\rm b}$ in power 
on all the scales where the NG components dominate,
the NG features of $\Delta_{\rm a}$ and $\Delta_{\rm b}$ 
are still fully recovered in $\Delta^{\rm W}$.
This is essentially because 
the intrinsic NG features of
these NG components are uncorrelated.

We also tested our method against the CLT,
which many statistics in the literature suffer from,
especially for those designed in the Fourier space
(e.g.,~\cite{8}).
To do this,
we considered the linear sum $\Delta=\Delta_{\rm (G)}+\Delta_{\rm (NG)}$
of a Gaussian background $\Delta_{\rm (G)}$ and
200 randomly distributed filled rectangles $\Delta_{\rm (NG)}$
(or diffuse points),
with an RMS ratio $\sigma_{\rm (G)}/\sigma_{\rm (NG)}=10^6$.
In the resulting $\Delta^{\rm W}$,
the shapes of the 200 rectangles 
(or the centers of the diffuse points)
are clearly recovered.
Thus we have seen a treatment which combines both the real and Fourier spaces
to take the advantages of both, 
without being affected by most of their disadvantages.

\begin{figure}[t]
  \centering
  \leavevmode\epsfxsize=8cm \epsfbox{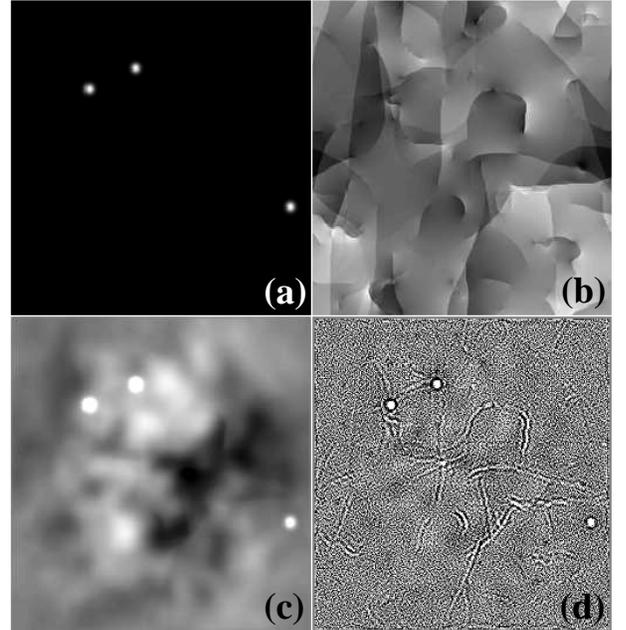}\\
  \caption
  {
    Simulated CMB maps of $2^\circ \times 2^\circ$:
    (a) point sources $\Delta_{\rm pnt}$;
    (b) cosmic-string-induced CMB $\Delta_{\cal S \rm ISW}$;
    (c) the total field $\Delta$;
    (d) the extracted NG signal $\Delta^{\rm W}$.
    The color schemes here are within the three-sigma range.
    }
  \label{fig-str}
\end{figure}

\noindent
{\bf D.\ Searching for cosmic strings}:
We first simulate a CMB field of $2^\circ\times 2^\circ$ 
with a resolution of $256^2$,
as observed from an interferometric experiment:
\begin{equation}
  \Delta
  =
  \left[
    (\Delta_{\rm bg}+\Delta_{\rm pnt}+\Delta_{\cal S \rm ISW})
    W_{\rm p}
  \right]
  \otimes
  W_{\rm o}+\Delta_{\rm noi}.
\end{equation}
Here
$\Delta_{\rm bg}$ is the Gaussian background 
(like the $\Delta_{\rm e}$ in the previous section),
$\Delta_{\rm pnt}$ is the point source (Fig.~\ref{fig-str}(a),
like the $\Delta_{\rm c}$ in the previous section but with $r=0.01$ here),
and 
$\Delta_{\cal S \rm ISW}$
is the string-induced CMB (Fig.~\ref{fig-str}(b)).
The $\Delta_{\cal S \rm ISW}$ is obtained using a toy model~\cite{Wu2001,myPhD}
which incorporates most important properties of cosmic strings
such as self-avoiding and scaling.
$\Delta_{\rm bg}$, $\Delta_{\rm pnt}$ and $\Delta_{\cal S \rm ISW}$
have RMS ratios $5:2:1$.
$W_{i}$ with $i=$ `p' and `o' 
denote respectively the primary and observing windows of a Gaussian form,
with Full Widths at Half Maximum of $0.7$ and $0.02$
in units of the field size.
$\Delta_{\rm noi}$ is a 5\% noise.

\begin{figure}[t]
  \centering
  \leavevmode\epsfxsize=8cm \epsfbox{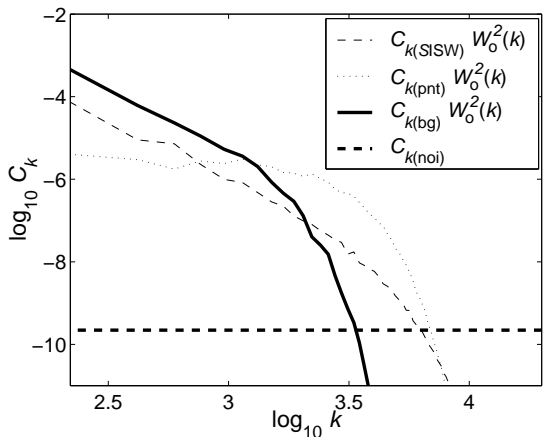}\\
  \caption
  {
    Power spectra of various components in Fig.~\ref{fig-str} (c).
    The $W_{\rm o}^2(k)$ is the power spectrum of the observing window.
    }                 
  \label{fig-str-pow}
\end{figure}

The resulting $\Delta$
and the whitened field $\Delta^{\rm W}$
are shown in Fig.~\ref{fig-str}
as (c) and (d) respectively.
Fig.~\ref{fig-str-pow} shows the power spectra of various components
in $\Delta$,
with
the convolution effect from $W_{\rm o}$ included.
As we can see,
although
the $\Delta_{\cal S \rm ISW}$ is hardly recognizable in $\Delta$
and its power (the dashed line in Fig.~\ref{fig-str-pow})
is dominated by other components on all scales,
its NG feature (the filament structure) 
is still fully recovered in $\Delta^{\rm W}$,
even with the fact that
the amplitude of the $\Delta_{\rm pnt}$ in $\Delta^{\rm W}$
is eighty times the RMS of $\Delta^{\rm W}$.
We have also applied our new method to
simulations of single-dish experiments,
where
$\Delta =
  \left[ (\Delta_{\rm bg}+\Delta_{\rm pnt}+\Delta_{\cal S \rm ISW})
  \otimes W_{\rm o} \right] W_{\rm p}+\Delta_{\rm noi}$.
The main conclusions remain unchanged.
In general we expect the string-induced NG feature to show up 
towards smaller scales, 
if not severely obscured by the noise.
This is because 
the $C_{k\rm (bg)}$ must have an exponentially decaying tail
due to the photon diffusion damping \cite{Silk}
while the $C_{k(\cal S \rm ISW)}$ must have a power-law decay 
due to its step-function like anisotropies on small scales.

\noindent
{\bf E.\ Discussion and conclusion}:
We proposed a new method of extracting the NG features of a given field.
Using the CMB anisotropies as an example, 
we numerically justified this method under various challenging circumstances.
In particular we show its capability of detecting cosmic strings.
With even more tests that are similar to those implemented above,
the main observation remains the same:
in a field $\Delta=\Delta_{\rm (G)}+\Delta_{\rm (NG)}$,
regardless how stronger the RMS of $\Delta_{\rm (G)}$ is,
the NG features of $\Delta_{\rm (NG)}$ can always show up 
in the whitened field $\Delta^{\rm W}$
as long as
$C_{k{\rm (NG)}}$ dominates $C_{k{\rm (G)}}$
within a certain range of $k$,
even if it is only a tiny fraction of the total accessible $k$ range.
In the extracted NG field $\Delta^{\rm W}$,
the NG features of uncorrelated NG components do not mix up
even if some of them dominate the others in power.
We also found that 
the new method is insensitive to the CLT.
With such whitened fields,
we can then further apply conventional statistics
to, for example, locate the NG scale.
Although some parameters in our simulations may not be realistic,
we chose these values only for the convenience of demonstrating
the power of our method under various situations.
The generalization of this method for the treatment of a full spherical CMB sky
is straightforward and
will be investigated in more details elsewhere \cite{Wu2001}.
It should be also noted that
our formalism is similar to the pre-whitening process commonly 
used in various context of data analysis,
but applied as a tool for extracting NG features for the first time here.

We also notice that
according to equation (\ref{Delta_p}),
in principle we can design a `window function' $P_k$
to remove the power on scales where the Gaussian signals of a field dominate,
so that 
the resulting $\Delta_{P}$ will have very well amplified NG feature.
However,
in general we do not know on which scale the NG signal dominates.
Thus
the method proposed here,
which effectively equalizes the power on all scales,
is optimal when no priori knowledge about a given field is available.
This may even enable us to find the NG signals of unknown physical processes.
The use of the $P_k$ in our formalism for various purposes 
will be further investigated elsewhere \cite{Wu2001}.
In conclusion,
we expect the new formalism to be
useful not only for the CMB observations in the near future,
but also for other fields of science
where identifying the NG components in a given field is important.

We thank 
Andrew Liddle, David Spergel, Neil Turok and Serge Winitzki
for useful conversations.
We are especially indebted to Paul Shellard,
under whose supervision
the main part of this work was finished \cite{myPhD}.
We also acknowledge the support from 
NSF KDI Grant (9872979) and
NASA LTSA Grant (NAG5-6552).



\end{document}